\documentclass[9pt,twocolumn,twoside]{opticajnl}
\journal{opticajournal} 

\setboolean{shortarticle}{true}

\usepackage{amsmath}
\usepackage{bm}
\usepackage{lineno}

\title{Direct and mediated dipole-dipole interactions in a reconfigurable array of optical traps}

\author[1]{Mian Wu}
\author[1,2]{Nan Li}
\author[1,3]{Han Cai}
\author[1]{Cheng Liu}
\author[1,*]{Huizhu Hu}

\affil[1]{State Key Laboratory of Modern Optical Instrumentation, College of Optical Science and Engineering, Zhejiang University, Hangzhou 310027, China}

\affil[2]{nanli@zju.edu.cn}
\affil[3]{hancai@zju.edu.cn}
\affil[*]{huhuizhu2000@zju.edu.cn}

\begin{abstract}
Optically levitated nanoparticles in vacuum experience both electrostatic and light-induced dipole–dipole interactions, offering a versatile platform to explore mesoscopic entanglement and many-body dynamics. A significant challenge in optical trap arrays is to achieve \emph{site-resolved, point-to-point} tunability: adjusting the laser parameters of a single trap typically induces global cross-talk to neighboring sites, hindering independent control. Inspired by tunable couplers in superconducting circuits, we implement an ancillary nanoparticle that functions as a coupler between two target nanoparticles. Within a reconfigurable three-particle array, we demonstrate broad tunability of the direct dipole–dipole interaction by controlling the phase and position of the traps. In addition, we observe spectral signatures consistent with mediated interactions between the target particles via the ancillary one, manifested as mode participation beyond the uncoupled response. Our results establish a practical route to tailored, site-resolved control in multi-particle optical trap arrays, expanding the optical-binding toolbox and opening opportunities for programmable oscillator networks relevant to macroscopic quantum mechanics and precision sensing.
\end{abstract}

\setboolean{displaycopyright}{false} 

\begin{document}

\maketitle

\section{Introduction}

Optically levitated mesoscopic particles \cite{Ashkin1971,Ashkin1976,Ashkin1986} provide an exceptionally clean platform with minimal mechanical contact, enabling ultra-sensitive force detection \cite{Butts2008,Monteiro2017,Moore2014,Ranjit2016,Hoang2016,Zhu2023} and access to the quantum ground state of motion in vacuum \cite{Barker2010,Romero-Isart2010,Chang2010,Delic2020,Piotrowski2023,Ashkin1977,Tebbenjohanns2021,Kamba2022,Yin2013,Chauhan2020,Arndt2014,Rudolph2022}. Leveraging this isolation, a single optically trapped nanoparticle has served as a testbed for precision metrology and stochastic dynamics. To probe quantum decoherence mechanisms, non-equilibrium phenomena, and emergent collective behavior, however, one must move beyond single traps toward arrays with controllable couplings \cite{Gonzalez2020}.

\begin{figure}[ht]
\centering
\fbox{\includegraphics[width=0.96\linewidth]{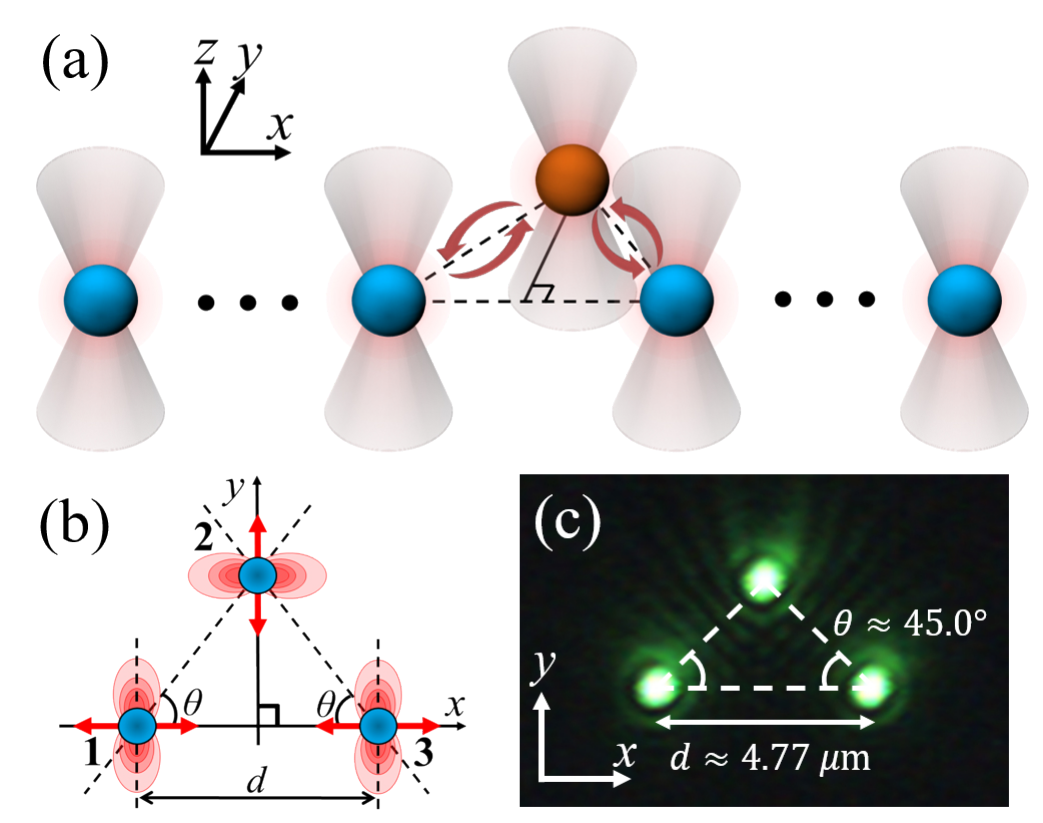}}
\caption{(a) Schematic of an ancillary nanoparticle used as a tunable coupler between two target nanoparticles in an optical trap array. Blue spheres denote the target array, and the orange sphere denotes the ancillary coupler. Red arrows indicate light-induced dipole–dipole interactions.
(b) Focal-plane arrangement: particles 1, 2, and 3 are trapped in traps 1, 2, and 3, respectively. Particles 1 and 3 are targets; particle 2 is the ancillary coupler. The traps form an isosceles triangle with base angles $\theta$ and base length $d$. The laser polarization of traps 1 and 3 is along $x$, whereas trap 2 is mainly polarized along $y$. Characteristic far-field dipole-radiation patterns under these conditions are also illustrated.
(c) Camera image of the three nanoparticles trapped under 532 nm illumination ($\theta \approx 45^{\circ}$, $d \approx 4.77\ \mu\mathrm{m}$).}
\label{fig:1}
\end{figure}
Recent experiments have shown that optical trap arrays support tunable, light-induced dipole–dipole interactions \cite{Rieser2022}, including non-reciprocal couplings that give rise to non-Hermitian or non-equilibrium  dynamics such as limit cycles \cite{Reisenbauer2024,Liska2024} and topological phases \cite{Hodaei2017,Fruchart2021,Loos2020,Liu2020,Xu2024,Youssefi2022}. Cavity fields can further mediate long-range interactions, with prospects for entanglement of mechanical modes \cite{Vijayan2024}. Yet a significant challenge remains to tune coupling in optical trap arrays without parasitic effects. Specifically, the coupling between two nanoparticles depends on the polarization, intensity, spacing, and phase of the trapping lasers at both ends \cite{Rieser2022}. Adjusting the laser parameters at any given site inevitably alters the coupling coefficients with all neighboring particles, degrading site selectivity. A proven solution in superconducting circuits is to interpose a tunable coupler between qubits \cite{Yan2018,Roushan2017,Deng2022}, which decouples local control from global cross-talk and underpinned landmark processors \cite{arute2019}.

Here we translate this concept to levitated optomechanics: we insert an ancillary nanoparticle as a coupler between two targets in a three-trap arrangement. By adjusting only the ancillary trap phase or position, we tune its direct coupling to each target while leaving the targets’ trap parameters fixed. We show wide-range control of direct couplings and observe signatures consistent with a coupler-mediated interaction between the targets. This approach adds a missing degree of freedom in controlling the optical trap arrays without resorting to globally invasive parameter changes.

\section{Theoretical model}
We consider three silica nanoparticles of radius $\sim 100$ nm levitated by 1064 nm light. In this Rayleigh regime, each particle acts as an induced dipole driven by the sum of its trapping field and the fields scattered by the others. The interference of these fields underlies light-induced dipole–dipole interactions. We focus on center-of-mass motion (COM) along the optical axis ($z$), as the coupling rates along the $z$-axis are significantly higher than those along the $x$- or $y$-axis\cite{Rieser2022}.

In our configuration, as shown in Fig.~\ref{fig:1}(b), traps 1 and 3 (the targets) lie on the $x$-axis, separated by a distance $d$, and trap 2 (the ancilla) lies on the perpendicular bisector, forming an isosceles triangle with base angles $\theta$; the traps are generated by light polarized along $x$ and $y$, respectively. In this geometry, the far-field radiation patterns suppress direct target–target coupling, while each target couples to the ancilla.
The non-reciprocal coupling coefficients between $z_1$ and $z_2$ can be assumed as $K\pm K'$ (plus sign for trap 2 to 1, vice versa) \cite{Rieser2022}. Here $K$ ($K'$) represents the (non-)conservative part of the interaction, respectively, given as
\begin{align}
K &= G\sin(2\theta) \cos\left( \frac{kd}{2\cos\theta} \right) \cos(\Delta\phi) \bigg/ \left( \frac{kd}{2\cos\theta} \right),  \label{eq:1} \\
K' &= G\sin(2\theta) \sin\left( \frac{kd}{2\cos\theta} \right) \sin(\Delta\phi) \bigg/ \left( \frac{kd}{2\cos\theta} \right),\label{eq:2}
\end{align}

where $G>0$ is a constant scaling with trap power and the particle polarizability, $k$ is the wave number of the trapping light, and $\Delta \phi=\phi_1-\phi_2$ represents the initial phase difference between traps 1 and 2 in the focal plane (the detailed derivation is presented in Section 1 of the Supplement 1). With the ancilla equidistant from the two targets and a $\pi$ phase offset between traps 1 and 3, the coupling coefficient between $z_3$ and $z_2$ is $-(K\pm K')$.

The particle COM positions $z_j$ ($j=1,2,3$) are governed by the linearized dynamic equations,
\begin{align} 
m\ddot{z}_{1} + m\gamma\dot{z}_{1} &= -\left(m\Omega_t^{2} + K + K'\right)z_{1} +\left( K + K'\right)z_{2}, \label{eq:3}\\
m\ddot{z}_{2}+m\gamma\dot{z}_{2}&=-m\Omega_a^{2}z_{2}+\left( K -K'\right)z_{1}-\left( K - K'\right)z_{3}, \label{eq:4}\\
m\ddot{z}_{3}+m\gamma\dot{z}_{3}&=-\left(m\Omega_t^{2} - K - K'\right)z_{3}-\left( K + K'\right)z_{2},\label{eq:5}
\end{align}
where the three particles share the same mass $m$, $\gamma$ represents the damping rate, corresponding to the mechanical linewidth, $\Omega_t = {\Omega}\sqrt{1+\eta}$ and $\Omega_a = {\Omega}\sqrt{1-2\eta}$ denote the intrinsic (uncoupled) mechanical frequencies for the target and ancillary traps, respectively. Here $\eta$ parametrizes the controlled power imbalance ($\Omega_i\propto\sqrt{P_i}$, $i=t,a$). For simplicity, the fluctuating forces arising from Brownian motion and other noise are not considered in the linearized dynamic equations. We compute the normal-mode frequencies $\omega_{1,2,3}$ numerically as functions of $d$, $\theta$, and $\Delta\phi$, and use them below to design and interpret measurements of direct and mediated interactions.

\section{Experimental setup}

\begin{figure}[ht]
\centering
\fbox{\includegraphics[width=0.96\linewidth]{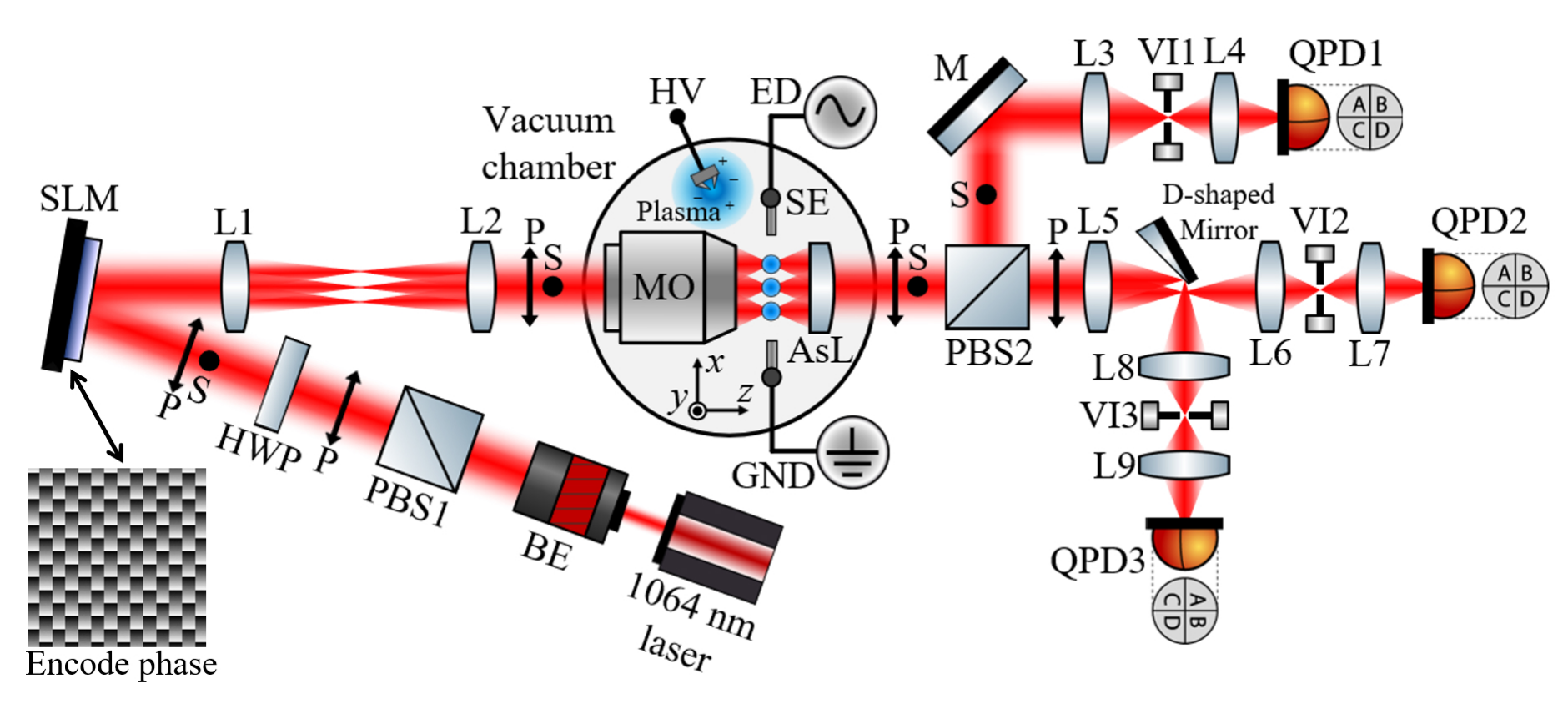}}
\caption{Schematic of the experimental setup. A 1064 nm trapping beam is expanded by a beam expander (BE), then passes through a polarizing beam splitter (PBS1) and a half-wave plate (HWP) to yield a linearly polarized beam with controllable P- and S-polarized components. The beam is incident on a spatial light modulator (SLM), using phase modulation to generate three optical traps. The forward-scattered light from each of the three traps is separated and then detected by quadrant photodiodes (QPD1–QPD3) for simultaneous motion readout. A high-voltage DC supply (HV) connected to bare wires inside the vacuum chamber is used to neutralize particle charge via a corona discharge, while a pair of steel electrodes (SE) applies an electric drive (ED) for charge monitoring. The setup also includes a multi-particle microscopic imaging system with 532 nm laser illumination, omitted from the schematic for clarity. L, lens; MO, microscope objective; AsL, aspheric lens; M, mirror; VI, variable iris.}
\label{fig:2}
\end{figure}

In our experiment, the trapping beam (ALS-IR-1064, Azur Light Systems, $\lambda=1064\ \rm{nm}$) was expanded to overfill the aperture requirements of the spatial light modulator (SLM, PLUTO-2-NIR-149, HOLOEYE) and the microscope objective (MO, TU Plan ELWD, Nikon, 100×, NA = 0.8). The beam, modulated by the SLM with an encode phase, passed through a 4f configuration composed of two lenses (L1 and L2, focal length 300 mm each). The encode phase, consisting of a binary phase grating with an added deflection phase, was imaged onto the front focal plane of the MO by the 4f configuration. The phase modulation of the SLM affected only light polarized along its long display axis (the P-polarized component, as shown in Fig.~\ref{fig:2}). Consequently, the P-polarized component of the beam formed traps 1 and 3 with nearly equal power, whereas the unmodulated S-polarized component formed trap 2. Rotation of the half-wave plate (HWP) tuned the power ratio $P_t/P_a$, while the encode phase on the SLM independently set trap positions and relative phases. The total power before the chamber was $\sim$2.5 W.

Silica nanoparticles (nominal radius $r_p=100$ nm) dispersed in isopropanol were introduced via ultrasonic nebulization. After loading single particles into each trap, the chamber was pumped to $\sim$10 mbar to maintain thermal equilibrium. To mitigate electrostatic interactions, a high-voltage DC supply was used to generate plasma containing positive and negative ions, thereby altering the net charge on the particles. Meanwhile, a pair of steel electrodes aligned along the $x$-axis applied an electric driving signal to the charged particles. The oscillation amplitude and phase of the particles under this driving signal were demodulated, allowing us to monitor and verify the neutralization of the net charge and the suppression of electrostatic interactions between the particles (see Section 2 of the Supplement 1 for details). An aspheric lens (AL1512-C, Thorlabs, NA = 0.55) was used to collimate the forward-scattered beam. The collimated beam then passed through a polarizing beam splitter (PBS2), where the beam of trap 2 was reflected, while the beams of traps 1 and 3 remained in the transmitted path. The transmitted beams of traps 1 and 3 were subsequently separated by a D-shaped mirror. Each separated beam was refocused and passed through a variable iris (VI) positioned at the focus to filter out stray light, thereby reducing signal crosstalk between different optical traps. The signals of each particle were detected by separate quadrant photodetectors (QPDs) and recorded synchronously. The $x$-mode signals were used to monitor the net charge of the particles, while the $z$-mode signals were converted into power spectral densities (PSDs) to characterize the interactions between the particles.

\section{Direct interactions in reconfigurable array}

\begin{figure}[ht]
\centering
\fbox{\includegraphics[width=0.96\linewidth]{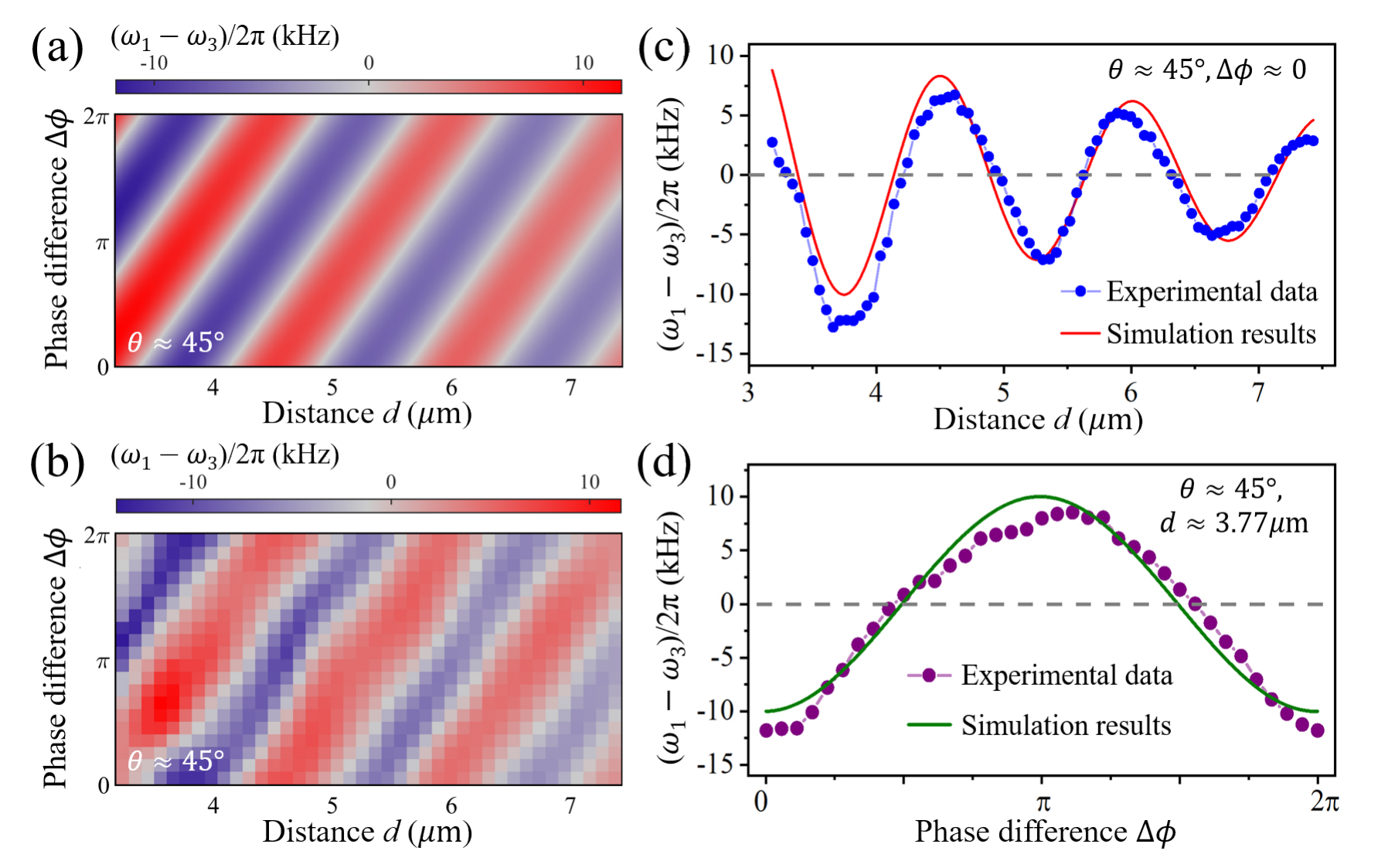}}
\caption{
(a) Numerically simulated map of the eigenfrequency splitting $(\omega_1-\omega_3)/2\pi$ as a function of  interparticle distance $d$ and phase difference $\Delta\phi$ at $\theta\approx 45^{\circ}$ and $\eta\approx-0.4$.
(b) Experimentally measured eigenfrequency splitting under the same conditions, showing periodic modulation with $d$ and $\Delta\phi$, in good agreement with simulation.
(c) Eigenfrequency splitting versus $d$ at near-zero phase difference $\Delta\phi\approx 0$. Blue dots: experimental data; red curve: simulation. Both show an oscillatory decay with period $\sim\sqrt{2}\lambda$.
(d) Eigenfrequency splitting versus phase difference $\Delta\phi$ at fixed $d=3.77$ $\mu\mathrm{m}$. Purple dots: experimental data; green curve: simulation. The modulation follows the interference phase.}
\label{fig:3}
\end{figure}

In the weak-coupling regime, perturbation theory implies that each normal mode—$\omega_1$ (target 1), $\omega_2$ (ancilla), and $\omega_3$ (target 3)—is a small shift from the corresponding intrinsic oscillator frequency. Because the frequency shifts of $\omega_1$ and $\omega_3$ depend on the respective ancilla–target coupling coefficients with opposite signs, we quantify the effective asymmetry by the difference between the two target-dominated modes, $(\omega_1 - \omega_3)/2\pi$, which serves as an experimentally convenient observable. Such a frequency difference tracks how the direct interaction varies under array reconfiguration, specifically with the phase difference $\Delta\phi$ and the geometric parameters of the isosceles arrangement, the target separation $d$ and base angle $\theta$.

With $\theta\approx45^{\circ}$ and $\eta\approx-0.4$, we map the frequency difference as a function of interparticle spacing $d$ and phase offset $\Delta\phi$. In Fig.~\ref{fig:3}(b), the measured data follows the calculated pattern in Fig.~\ref{fig:3}(a), showing a periodic modulation of $(\omega_1 - \omega_3)/2\pi$ as both $d$ and $\Delta\phi$ are varied. For $\Delta\phi=0$, the coupling coefficient $K$ decays approximately as $1/d$ while oscillating due to the dipole scattering in the far field, yielding an oscillatory $(\omega_1 - \omega_3)/2\pi$ with period $\sim\sqrt{2}\lambda$ at $\theta=45^{\circ}$ [Fig.~\ref{fig:3}(c)]. At a fixed $d\approx3.77$ $\mu$m, sweeping $\Delta\phi$ from $0$ to $2\pi$ produces a sinusoidal frequency difference, as depicted in Fig.~\ref{fig:3}(d), maximized near $\Delta\phi=n\pi$ (where $n$ is an integer), as expected from $K(\Delta\phi)$. Overall, the measurements and simulations agree closely, with small discrepancies attributable to slight trapping power mismatch and residual polarization imperfections.

\begin{figure}[ht]
\centering
\fbox{\includegraphics[width=0.96\linewidth]{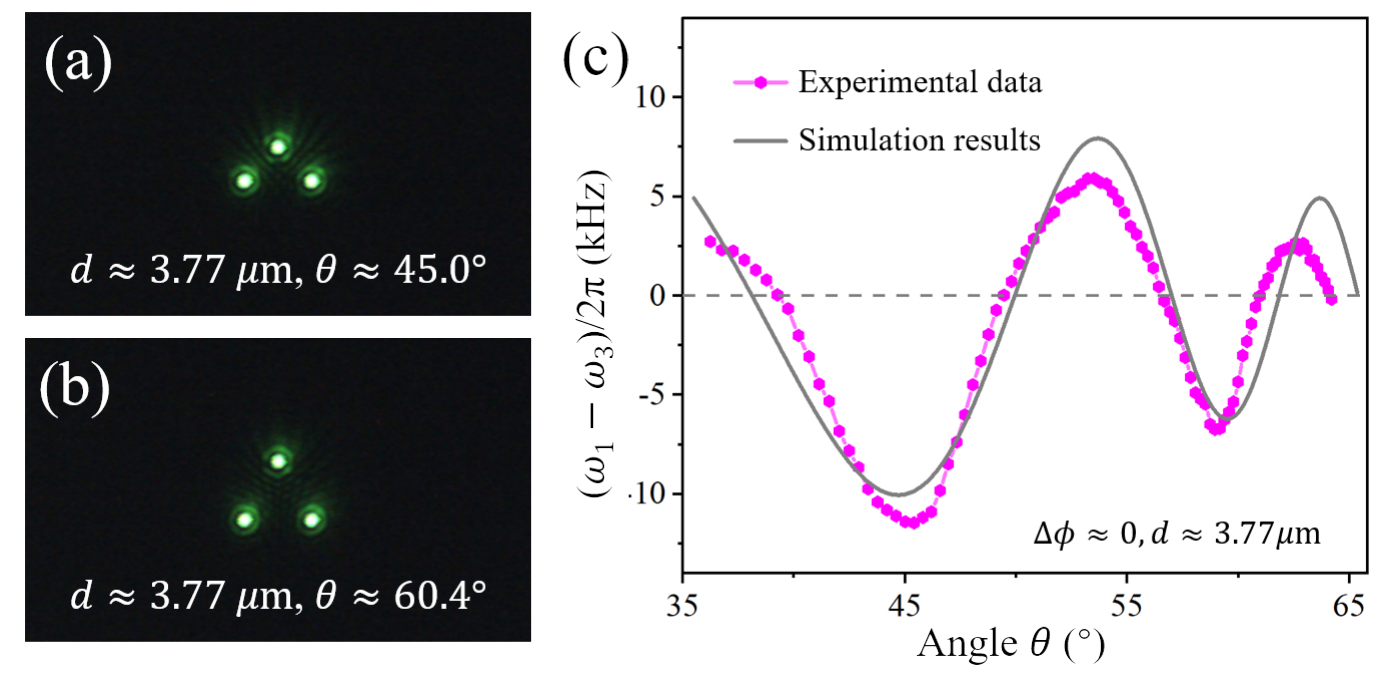}}
\caption{
(a) Camera image of three particles arranged in an isosceles triangle ($d\approx3.77$ $\mu\mathrm{m}$, $\theta\approx45.0^{\circ}$).
(b) Same distance $d$ but larger angle ($\theta\approx60.4^{\circ}$).
(c) Eigenfrequency splitting $(\omega_1-\omega_3)/2\pi$ versus $\theta$ for $d\approx3.77$ $\mu\mathrm{m}$, $\eta\approx-0.4$, and $\Delta\phi=0$. Magenta dots: experiment; gray curve: simulation. As $\theta$ increases, the splitting oscillates with gradually decreasing amplitude and period.}
\label{fig:4}
\end{figure}

Beyond phase $\Delta\phi$ and spacing $d$, moving only the ancilla changes the base angle $\theta$ and thereby the overlap of radiation patterns. Holding $d\approx3.77$ $\mu$m, $\eta\approx-0.4$, and $\Delta\phi=0$ fixed, we observe an oscillatory $(\omega_1 - \omega_3)/2\pi$ in the experimental data by varying the angle $\theta$ from $36.2^{\circ}$ to $64.2^{\circ}$ [Fig.~\ref{fig:4}(c)]. The data exhibits a pronounced extremum near $\theta\approx45^{\circ}$, a feature which is in close agreement with the theoretical prediction, and a gradual reduction in amplitude as $\theta$ increases. Deviations for $\theta\gtrsim60^{\circ}$ likely stem from radiation-pressure-induced micro-shifts of trap positions and a small residual P-polarized component in the ancilla beam due to $\sim 93\%$ SLM diffraction efficiency.

\section{Evidence of mediated interactions}

\begin{figure}[ht]
\centering
\fbox{\includegraphics[width=0.96\linewidth]{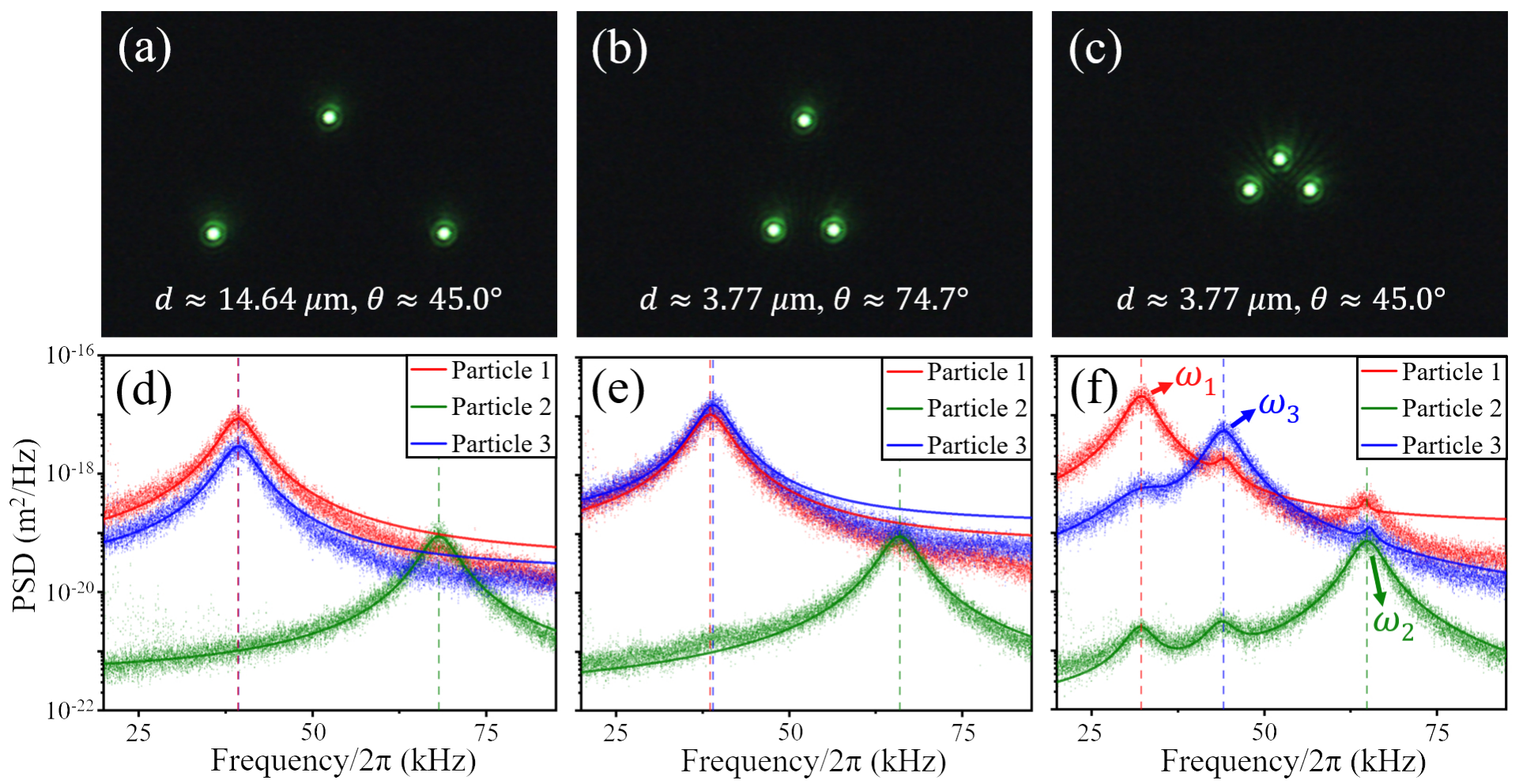}}
\caption{
(a–c) Camera images of three nanoparticles in different geometric configurations defined by $d$ and $\theta$.
(d–f) Corresponding power spectral densities (PSDs) of $z$-axis motion at $\eta\approx-0.4$ and $\Delta\phi=0$. For large separations (a,b), the PSDs (d,e) show distinct peaks at the intrinsic mechanical frequencies of each particle, indicating negligible coupling. When the particles are brought into close proximity (c), the ancillary particle interacts with the targets to form a coupled mechanical system. The PSDs (f) reveal three normal modes with eigenfrequencies $\omega_1$, $\omega_2$, and $\omega_3$, with the targets exhibiting participation at multiple mode frequencies.}
\label{fig:5}
\end{figure}

We next focus on whether the ancilla provides an effective target–target interaction. When all three particles are far apart, each PSD shows a single peak at its intrinsic frequency [Figs.~\ref{fig:5}(a,d)], indicating negligible coupling and independent harmonic oscillations. Bringing the targets closer while keeping the ancilla distant yields a decoupled response [Figs.~\ref{fig:5}(b,e)], consistent with the geometric suppression of the direct target–target channel in our configuration.

When the ancilla is brought into proximity with both targets, as shown in Fig.~\ref{fig:5}(c), the three-particle system exhibits three distinct normal modes in the PSDs [Fig.~\ref{fig:5}(f)]. Importantly, the target PSDs show participation at multiple mode frequencies, i.e., beyond their uncoupled responses. This redistribution of spectral weight is consistent with a second-order, ancilla-mediated coupling between the targets predicted by Eqs.~(\ref{eq:3}--\ref{eq:5}). Given the relatively weak far-field dipolar interaction strengths in our present parameter regime, we refrain from quantifying a tunable mediated coupling (e.g., via an independently calibrated splitting) and instead report spectral evidence, i.e., mode mixing and participation, that follows the qualitative trends of the model. Increasing the coupling rates (e.g., via higher power, optimized polarization purity, or cavity enhancement) should enable a more direct spectroscopic extraction of the effective mediated interaction.

\section{Conclusion}

We have implemented the tunable-coupler paradigm in a levitated optomechanical array by inserting an ancillary nanoparticle between two target ones. This architecture circumvents global cross-talk: by adjusting only the phase and position of the ancilla, we achieve wide-range control of the direct ancilla–target dipole–dipole interactions while keeping the target traps fixed. We also observe \emph{evidence} consistent with an ancilla-mediated interaction between the targets in the form of mode participation beyond the uncoupled response. The approach adds a powerful degree of freedom, namely, site-resolved, point-to-point tunability, to optical-binding platforms, and provides a scalable route toward programmable oscillator networks for macroscopic quantum mechanics, precision sensing \cite{McDonald2020}, and stochastic thermodynamics \cite{Gieseler2018,Fruchart2021,Mahault2022,Tang2021}. Future work will focus on amplifying and resolving the mediated channel to enable quantitative, tunable coupler operation and on extending the scheme to larger arrays.

\begin{backmatter}






\bmsection{Disclosures} The authors declare no conflicts of interest.

\bmsection{Data Availability Statement}  Data underlying the results presented in this paper are not publicly available at this time but may be obtained from the authors upon reasonable request.

\bmsection{Supplemental document}
See Supplement 1 for supporting content.
\end{backmatter}




\bibliography{Updated_Reference}

\bibliographyfullrefs{Updated_Reference}


\ifthenelse{\equal{\journalref}{aop}}{%

}{}

\end{document}